\documentclass[conference]{IEEEtran}
\IEEEoverridecommandlockouts
\usepackage[utf8]{inputenc}

\usepackage{cite}
\usepackage{amsmath,amssymb,amsfonts}
\usepackage{algorithmic}
\usepackage{graphicx}
\usepackage{textcomp}
\usepackage{xcolor}

\begin{document}
\pagestyle{plain}
\def\BibTeX{{\rm B\kern-.05em{\sc i\kern-.025em b}\kern-.08em
    T\kern-.1667em\lower.7ex\hbox{E}\kern-.125emX}}
\title{Towards an Interface Description Template for Reusing AI-enabled Systems\\
}

\author{\IEEEauthorblockN{Niloofar Shadab}
\IEEEauthorblockA{\textit{Grado Dept of Industrial and Systems Engineering} \\
\textit{Virginia Tech}\\
Blacksburg, VA, USA \\
nshadab@vt.edu}
\and
\IEEEauthorblockN{Alejandro Salado}
\IEEEauthorblockA{\textit{Grado Dept of Industrial and Systems Engineering} \\
\textit{Virginia Tech}\\
Blacksburg, VA, USA \\
asalado@vt.edu}

}

\maketitle

\begin{abstract}
Reuse is a common system architecture approach that seeks to instantiate a system architecture with existing components. However, reusing components with AI capabilities might introduce new risks as there is currently no framework that guides the selection of necessary information to assess their portability to operate in a system different than the one for which the component was originally purposed. We know from SW-intensive systems that AI algorithms are generally fragile and behave unexpectedly to changes in context and boundary conditions. The question we address in this paper is, what type of information should be captured in the Interface Control Document (ICD) of an AI-enabled system or component to assess its the compatibility with a system for which it was not designed originally. We present ongoing work on establishing an interface description template that captures the main information of an AI-enabled component to facilitate its adequate reuse across different systems and operational contexts. Our work is inspired by Google's Model Card concept, which was developed with the same goal but focused on the reusability of AI algorithms. We extend that concept to address system-level autonomy capabilities of AI-enabled cyberphysical systems.
\end{abstract}


\section{Introduction}
Reuse is a common system architecture approach that seeks to instantiate a system architecture with existing components instead of developing dedicated ones with the objective to reduce the costs and risks associated with developing a new product\cite{beam}\cite{ontology}. Vendors of these existing components, often called off-the-shelf (OTS) products or Non Developmental Items (NDI), provide characterization information that systems engineers use to inform requirements derivation, architectural choices, and test cases\cite{OTS}\cite{maturity}. The amount of information that is provided with the component varies depending on the specific application domain and project, spanning from a simple datasheet to a data package that contains detailed design and certification information\cite{dod}. However, reusing components with AI capabilities might introduce new risks as there is currently no framework that guides the selection of necessary information to assess their portability to operate in a system different than the one for which the component was originally purposed.

For example, consider a component built of hardware (HW) and software (SW) that implements an AI capability and which was originally designed to operate as part of a certain system (and hence of operational environments). Assessing the adequacy of the component to be reused in a different system usually consists of evaluating the compatibility of its external interfaces, the compatibility of the component with the new environmental conditions, and the provision of the required functionality and performance. These compatibility assessments can often be done by looking at the component's Interface Control Document (ICD), which lists the energy resources that the component needs, the physical and functional properties of its interfaces, data structures and communication protocols, environmental conditions that the component can sustain and those that it generates, etc.\cite{nasa}. The question we address in this paper is, what type of information should be captured in such ICD to assess the compatibility of an AI-enabled component within a system for which it was not designed originally.

We know from SW-intensive systems that AI algorithms are generally fragile and behave unexpectedly to changes in context and boundary conditions \cite{robust}\cite{trust2}. Specifically, when used across different applications and with different datasets, AI algorithms may suffer from AI fairness and AI inclusion\cite{fair-def}\cite{report}. Understanding the development and training processes of these AI algorithms is essential to properly evaluate fragility\cite{fairness}\cite{NN}. As AI-enabled components become ubiquitous in many fields and applications\cite{app1}\cite{app2}, we suggest that it is necessary to understand the meaning, implications, and factors of fragility in the context of cyber-physical systems that embed AI capabilities throughout multiple integration levels that blend functionality in hardware and software.

In this paper, we present ongoing work on establishing an interface description template (or information framework) that captures the main information of an AI-enabled component to facilitate its adequate reuse across different systems and operational contexts. Our work is inspired by Google's Model Card concept\cite{card}, which was developed with the same goal but focused on the reusability of AI algorithms. We extend that concept to address system-level autonomy capabilities of CPS. In this paper, we aim to integrate interface descriptions used for traditional components with descriptions used in AI model cards to sufficiently characterize interface of AI-enabled components in order to provide a platform for assessing AI-enabled component's portability and reuse in different types of systems.

This paper is organized as follows. In section II, we briefly present some background of the concepts that we use in this paper, including common interface descriptions for hardware and software components. In section III, we present the proposed interface description template and discuss the challenges in using them within the context of AI-enabled CPS.

\section{Background}
In this paper, we define a CPS as a system that integrates SW and HW components to yield desired capabilities, where the functionality of the HW components extends beyond providing computational resources to run the SW\cite{cps-def}. By AI-enabled components or systems, we refer to those CPS that exhibit autonomy, collaboration, and adaptation\cite{complexity}. Here, we present prior efforts to characterize AI algorithms and the data they use for training. In addition, given the importance of the dynamic behavior of AI-enabled components to characterize their interface as part of a complex systems, we also discuss some of the properties of autonomy capabilities in AI-enabled CPS. Specifically, we focus our attention to adaptability, changeability, and cooperation.

\subsection{Model Reporting Used for AI Algorithms}
Model Cards have been proposed as a possible solution to problems of AI fairness\cite{card}. Specifically, it is argued that a Model Card enables transparent AI model reporting across different AI model users aiming at preventing unfair and inappropriate use of AI algorithms in contexts that are not suitable for their use\cite{card}. Model Cards are short documents accompanying trained machine learning models that provide benchmarked evaluation in a variety of conditions that are relevant to the intended application domains. A Model Card template is shown in Table \ref{table_1}.

\begin{table}[h]
\caption{Model Cards Documented Features}
\label{table_1}
\centering
\begin{tabular}{|p{0.43\textwidth}|}
\hline
Model Cards \\
\hline
\textbf{Model Details}. Basic information about the model such as\\
– Model date\\
– Model version\\
– Model type\\
– Information about training algorithms, parameters, fairness constraints or other applied approaches, and features\\
\hline
\textbf{Intended Use}. Use cases that were envisioned during development.\\
\hline
\textbf{Factors}. Factors could include demographic or phenotypic groups, environmental conditions, technical attributes\\
\hline
\textbf{Metrics}. Metrics should be chosen to reflect potential real-world
impacts of the model such as\\
-Model performance measures\\
– Decision thresholds\\
– Variation approaches\\
\hline
\textbf{Evaluation Data}. Details on the dataset(s) used for the
quantitative analyses in the card.\\
– Datasets\\
– Motivation\\
– Preprocessing\\
\hline
\textbf{Training Data}. Minimal allowable information
should be provided here, such as details of the distribution over various factors in the training datasets.\\
\hline
\textbf{Quantitative Analyses}\\
– Unitary results\\
– Intersectional results\\
\hline
\textbf{Ethical Considerations}\\
\hline
\textbf{Caveats and Recommendations}\\
\hline
\end{tabular}
\end{table}

The Model Card framework reports the intended purposes of developing an AI algorithm and the underlying characteristics of the data that were used to train and test these algorithms. It also reports the statistical properties of the data and the performance of the algorithm using metrics such as false-positive rates. By attaching the Model Card to the corresponding AI algorithm, it is fairly easy to reach out for such information to identify the possible biases in the training and testing of the algorithm when considering using the AI algorithm in applications or contexts for which it was not originally intended.

\subsection{Interface Description Templates}
ICDs are essential artifacts in systems engineering that contain critical (and often sufficient) information to assess the interoperability between components and hence the portability of components among different systems\cite{sebok}. In order to do so, interfaces are comprehensively described, including characterization of the logical signal conveyed through the interface and information about the interface at different layers of abstraction (e.g., physical layer, transport layer, etc. \cite{7layer}). 

The specific information contained in ICDs is often driven by the particular application domain. Generally, predefined templates exist for both HW-intensive systems and SW-intensive systems; although these are combined for CPS. 

In HW-intensive systems, the connection between two systems can be abstracted as a logical signal being conduced through a physical interface\cite{truembse}. The logical signal represents the element that is transferred between the two systems, whereas the physical interface represents the conduit that enables such transfer. Logical signals in this sense can be of the form information, energy, or material\cite{kossiakof}. Physical interfaces are more sophisticated and, as stated before, can be characterized in a number of layers. For simplicity purposes, we consider in this paper only a physical layer (the actual medium that will conduct the signal) and a transport layer (which codes information into a given representation). For example, a system can send its temperature (signal in the form of information with a given range of temperatures) as a series of bits, for which a mapping between series of bits and temperature is given, using a communication protocol (which indicates start of sequence, end of sequence, correction of errors, etc.), coding the bits as differences of electrical voltage, and sending those voltage differences through a given pair of wires, which are connected on specific pins of a connector, on which the wires exercise forces and conduct thermal fluxes. All of this happening in certain electromagnetic environment and subjected to certain temperatures (these are different than the information that is sent) and levels of humidity, among others. Note that, given the same information signal, the physical interface could have been different, such as through encoding the temperature as an analogue electrical signal or observing infrared emissions through the air. An interface description template for an interface would therefore include all such information. A partial example is given for illustrative purposes in Table \ref{table_6}.

In SW-intensive systems, the connection between two system can be abstracted as the information being transformed through data stream, pipelines and filters which are defined as interfaces\cite{cloud}. Interfaces in software system provide detailed information about the functionality and capabilities of the software component. To cover all the aspects of the software component functionality, one needs to specify properties, operations and events for the interface. Properties refer to the observable structures of the interface, operations represent the dynamic behavior of the software system with its environment in a proactive manner and event captures the event-based behavior of the software in a reactive manner\cite{sw-intensive}. The packaging and configuration information is needed for software component to be able to utilize them within systems and operating contexts. For example, when a component is supposed to work in a specific context, it operates and interacts with different components. These interactions between components and the perspectives with which these interacts happen may affect the set of visible properties as well as allowable operations and events. Another important aspect of the software interface is the non-functional properties such as the level of security and reliability of the interface to transfer information across components. Therefore, an interface description template for a software interface would include all such information. A partial example is given for illustrative purposes in Table \ref{table_7}. 
 
\begin{table}[h]
\caption{Summary of Interface Description Template for Hardware Component}
\label{table_6}
\centering
\begin{tabular}{|p{0.09\textwidth}|p{0.34\textwidth}|}
    \hline
    Dimension & Interface Attributes \\
    \hline
    Signals & \textbf{Type}:\\
    & Information, Energy (e.g., Electrical, Mechanical, Chemical, Nuclear, Radiant, Light, Sound), Material \\
    & \textbf{Characteristic 1 .. n}:\\
    & Properties of the signal that is being conveyed\\
    \hline
    Interface & \textbf{Physical Layer}: Each feature should be described for In and Out of the interface separately.\\
    & – \textit{Electrical Properties}:\\
    & EMC (Impedance, Radiated susceptibility, Conducted susceptibility, ESD susceptibility) , Communication (Voltage, Current, Frequency, Signal shape)\\
    & – \textit{Mechanical Properties}: \\
    & Random vibration levels, Acoustic vibration levels, Shock levels, Bending and other forces.\\
    & – \textit{Thermal Properties}:\\
    & Thermal greases, Phase change materials, Thermal tapes and Gap filling thermal pads.\\
    & – \textit{Particulate Properties}\\
    & \textbf{Transport Layer}: This layer codes information into a given representation\\
    \hline
\end{tabular}
\end{table}

\begin{table}
\caption{Summary of Interface Description Template for Software Component}
\label{table_7}
\centering
\begin{tabular}{|p{0.09\textwidth}|p{0.34\textwidth}|}
    \hline
    Dimension  & Interface Description \\
    \hline
    Signal & \textbf{Type}:\\
    & Information\\
    \hline
    Interface & \textbf{Properties}:\\
    & – Externally observable structural elements of the component. Some are just observable, the others can be observed and changed by the user.\\
    & \textbf{Operations}:\\
    & – Proactive control (dynamic behavior capabilities) with which the component interact with the environment. Provides service/functionality. \\
    & \textbf{Events}:\\
    & – Reactive control that the component may generate so that other components in the system might choose to respond to.\\
    & \textbf{Constraints}:\\
    & There are two types of constraints. One on individual elements of the component and the other is on the relationships among these elements.\\
    & \textbf{Packaging and Configuration}:\\
    & Packaging defines two aspects of the component. One is the role of the component operating in a context. The other is the set of contexts that the component may be used in.\\
    & \textbf{Non-functional Properties (ilities)}:\\
    & Such as security, performance, reliability, etc and how they are characterized and what are the impacts to the system having this component in different contexts.\\
    \hline
\end{tabular}
\end{table}

\section{Autonomy Properties of AI-enabled components}
We suggest that in order to provide a comprehensive description of the interface of the AI-enabled components, it is necessary to capture specific properties that autonomy brings to these systems. In this paper, we specifically focus on adaptability, changeability, and cooperation properties, and identify the important aspects of these properties to form the description of their interfaces.

\subsection{Adaptability}
A system is designed to operate in different environments and/or operational conditions and to interact with a range of other systems (collaborating systems or/and competing systems)\cite{samurai}, depending on its intended use. However, these may change while the system is operational\cite{framework} and an autonomous system is likely desired to adapt to those previously unknown operational conditions, environments, and external systems\cite{safety}\cite{uncertainty}. Therefore, we suggest that interfaces that enable the adaptation capability of the system need to capture the source of the changes and the system-level responses to adapt to such changes. We consider in this paper two types of changes. First, there is the possibility of change in the problem space for which the system was designed, that is, changes in stakeholder needs, operational conditions, external systems, etc. In this case, there is a question of whether the existing system still belongs to the solution space for the changed problem space, and if the answer is negative, how it can adapt to be the part of the new solution space. Second, there is the possibility of changes within the system itself, as a result, for example, of a need to update, upgrade, or maintain the system. In this case, the system faces similar consequences as in the first case, and the system might leave the solution space. In this situation, the system needs to adapt to come back to the solution space that addresses the original problem.

For the first scenario, the new environment (problem space) is unknown for the AI-enabled component within the system of interest. Thus, once in one of these unknown environments, the AI-enabled component needs to update and adapt its learning process (or knowledge representation) using the new information it receives from the interactions with the environment, which can be internal (through the transformation executed by other components in the system) through its interfaces. Accordingly, the AI component requires specific capabilities to guarantee the consistency of its adapted capabilities with those that the system needs to provide externally. In the second scenario, change may happen gradually, such as small but frequent software updates, but eventually, imply a significant departure from the initial design. As a result, the AI-enabled component also changes gradually by updating its learning process based on system-level changes.

In both scenarios, capturing AI adaptation capability in the interface description of AI-enabled component can be useful to prevent malfunctions in these systems when they are not compatible with their operating environment. This specifically becomes important because, instead of failing or degrading, an AI-enabled component might just continue to operate but with undesired/unintended outputs. This might provide a problem of late detection of failure compared to the traditional components that could signal malfunctions and have predictable failure rates.

Similar to SW-intensive AI systems\cite{catastrophic}\cite{robust}, adaptation of AI-enabled systems are likely to be vulnerable to catastrophic interference, where AI-enabled components forget about the past learning once they get new information. These issues bring up the problem of the reliability of autonomy capabilities of complex systems when they are exposed to new scenarios. This indicates that the set of possible environments the system might operate in, in the future, is an important factor to understand the compatibility of the AI-enabled components interface with the system of interest.

The compatibility of AI-enabled components with the environment that the system of interest operates in highly depends on both internal factors (such as the training distribution, the assumption of a priori, and the sensitivity level of the algorithm), as well as external interface factors (such as how other parts of the system filter/transform the environment into the specific interfaces of the AI-enabled component). As a result, these parameters should likely be included in the interface description templates for AI-enabled components.

\subsection{Changeability}
Changes in the system of interest can be characterized and quantified by three elements: agent of the change (the forces that trigger the change), the mechanisms of the change (the path that system takes to transition from its state to an altered state after change agent forces the system to change), and the effect of the change (the difference between the system's prior state and the altered state after the change mechanism happens)\cite{changeability}, all which may be uncertain, as shown in Table \ref{table_2}.
\begin{table}[h]
\renewcommand{\arraystretch}{1.3}
\caption{Sources of Uncertainty in System Changes}
\label{table_2}
\centering
\begin{tabular}{|p{0.07\textwidth}|p{0.1\textwidth}|p{0.1\textwidth}|p{0.11\textwidth}|}
\hline
Uncertainty & State of Change & Change Mechanism & Agent of Change \\
\hline
1 & Known & Known & Known \\
\hline
2 & Unknown & Known & Known\\
\hline
3 & known & Known & Unknown\\
\hline
4 & Known & Unknown & Known\\
\hline
5 & known & Unknown & Unknown\\
\hline
6 & Unknown & Unknown & Known\\
\hline
7 & Unknown & Known & Unknown\\
\hline
8 & Unknown & Unknown & Unknown\\
\hline
\end{tabular}
\end{table}

The nature of the changes the system goes through should hence be a factor in the changeability to be exhibited by the AI-enabled component. As changes take place, the system of interest might encode data differently from how it was designed initially. This means that the AI-enabled component, which was trained and worked with data encoded using the initial encoding approach, might be incompatible with the data produced by the system after the change occurs. Incompatibility, in this case, refers to the scenarios where interoperability of components in the system provide negative effects on the AI-enabled component operations due to the fragility of AI functions. Specifically, the AI component might work with the new encoded data apparently well, but actually in a way that might result in unexpected behaviors. This phenomenon in AI algorithms that can happen in a slowly emerging process is called "Drift of Concept"\cite{drift}. It is defined as changes that make the model built on old data inconsistent with the new data. The problem of concept drift becomes more important when the AI-enabled component is intended to be used in a CPS. This is due to the fact that tracking down the changes withing a CPS and tracing them back to the learning task of AI-enabled component could be challenging.

During the development of an AI-enabled CPS, information about the ability of the system to handle concept drift is necessary when evaluating the changeability capability of the system over its life-cycle. Moreover, due to insufficient data, some AI-enabled components may be unable to adapt to new elements that are substitute existing ones or added to the system. Instead, the adaptation of the AI-enabled component is delayed with respect to the contextual change, which interrupts the performance of the system in its new context. As a result, we believe that information on testing of AI-enabled component dealing with these changes can be a good indication of whether the AI-enabled component is suitable with the CPS changeability capability. The interface description template needs to take all such information into account to prevent inappropriate usage of AI-enabled components that cannot support the level of changeability of the AI-enabled CPS.

\subsection{Cooperation}
AI-enabled systems may engage in cooperation and/or interactions with other systems or humans. In this paper, we limit Cooperation to the capability of AI-enabled CPS to participate in 1) zero-sum scenarios, that is, scenarios in which cooperating systems have a common interest, and 2) scenarios in which human and/or the CPS preferences are neither fully aligned nor fully in conflict with each other\cite{colab}. Cooperation is needed when dealing with multi-agent AI-enabled CPS where multiple AI-enabled components need to cooperate with each other in a network within CPS. The degree of cooperation required for CPS operation can be determined by AI capability in support of the system and/or the user as well as the joint activities with other subsystems and/or external systems to achieve a long-term goal of the system\cite{coop}. As a result, during the system design phase, we need to understand the number of interfaces required for AI-subsystem to interact with in order to achieve a desired level of cooperation.

\section{Interface Description Template for AI-enabled components}
Table \ref{table_3} and Table \ref{table_8} list a summary of the features of the an AI-enabled components that we suggest should be included in an ICD. These features have been derived from the discussion presented in the previous section, using the Model Cards as a paradigm and inspired by the characterization of self-organizing complex system patterns\cite{self-org}\cite{decentralized}, which represent the autonomy aspect of this type of components. A description of such features (or attributes), together with those traditionally employed for HW systems (ref. Table \ref{table_6}) and for SW systems (ref. Table \ref{table_7}), form our proposed interface description template.  

The information that the interface description should capture is split in two categories. Table \ref{table_3} lists the information of the edge scenarios that the AI-enabled component may be exposed, which can be considered as extreme events that might interrupt or deviate the operations of the AI-enabled CPS from its main mission. These conditions are identified as: Catastrophic Interface, Drift of Concept, and cooperation deviation scenarios such as unintended competition and deviation from system goals. The other factors are related to system-level behavior and structure of the AI-enabled system: Decentralization, sub-optimality, and unintended synergy. A decentralized solution might increase the overall complexity of the design and the complexity of the interactions of AI-enabled components in scenarios such as network of autonomous agents. The concept of \textit{optimal solution} does not hold for AI-enable components given their evolution their life-cycle. Therefore, the trade-off between optimal solution and autonomy should be considered once utilizing AI-enabled components. Synergy is a byproduct of adaptability of AI-enabled CPS. As a result, the possibility of unintended synergy behaviors should be considered while working with AI-enabled components, especially with various level of required security and reliability within the system.

\begin{table}[h]
    \renewcommand{\arraystretch}{1.3}
    \caption{A summary of the considerations that need to be described in The Interface Description of AI-enabled Components}
    \label{table_3}
    \centering
    \begin{tabular}{|p{0.09\textwidth}|p{0.34\textwidth}|}
    \hline
 AI-enabled Component & Considerations Description\\
    \hline
    Considerations & - \textbf{Catastrophic Inference}: AI-enabled component might suffer from Catastrophic Inference. This causes the new set of data completely overtake the previous learning process of AI-enabled component.\\
    \hline
    & - \textbf{Drift of Concept}: The AI-enabled component might suffer from Drift of Concept where the model built on old data inconsistent with the new data of system. \\
    & - \textbf{Decentralization}: Changes in one part of the system can be a decentralized process; thus might need large amount of communication and interactions between different parts of the system with AI-enabled component; as a result increasing the complexity of the interconnections.\\
    & - \textbf{Optimality}: Using AI-enabled component needs trade-offs between optimality and flexibility of the design to respond to changes in problem space. Because optimum solution exists just in static situations. \\
    & - \textbf{Unintended synergy}: Information of the possible unintended synergy behavior.\\
    & - \textbf{Unintended Competition}: Information of the possible unintended competitive behavior.\\
    & - \textbf{Deviation from system goals}: information on the greedy approaches that can be problematic as the agents look for maximizing local rewards while ignoring long-term systemic-oriented goals. \\
    \hline
        \end{tabular}
\end{table}

Table \ref{table_8} describes the interfaces of an AI-enabled component that should be evaluated to assess the component's reusability in different systems and operating environment. As in SW-intensive systems, aspects of the interface related to the physical layers and observable structural elements are relevant for an AI-enabled component. The behavioral patterns of the AI-enabled component in its environment, through its interface (specifically the exploration and exploitation behaviors) give information on the expected patterns of interactions of the AI-enabled component with other components as well as with its the environment. The Degree of flexibility in the solution space, which is given by the capability of the AI algorithm implemented in the AI-enabled component to adapt to new situations as transformed through the HW elements, is also incorporated in the template. The Sensitivity level of the design space of the AI-enabled component provides an indication of the reliability of the AI-enabled component when dealing with the changes within the system. Operations and events take a similar form as those in SW-intensive systems. In this case, an example of an event-based actions can be the location-based data of the system so that the AI-enabled component needs to respond to accordingly. Information on local components interactions protocols and constraints on the human interactions are also reported. Note that in this template, reports on the verification strategies used to verify AI-enabled components are critical pieces of information, since the verification procedures can also play roles in the learning process of the AI-enabled component. Moreover, for AI-enabled components, different verification strategies verify different aspects of the AI-enabled behaviors with different degree of confidence, since the performance of the AI algorithm is coupled with the HW and SW elements that provide data to/from it.

\begin{table}[]
    \renewcommand{\arraystretch}{1.3}
    \caption{A summary of the features that need to be described in The Interface Description of AI-enabled Components}
    \label{table_8}
    \centering
    \begin{tabular}{|p{0.09\textwidth}|p{0.34\textwidth}|}
    \hline
    AI-enabled Component & Interface Description\\
    \hline
    Interface & - \textbf{Properties}: Physical layers and observable structural elements of the AI-enabled component.\\
    & - \textbf{Exploration vs Exploitation}: The mechanisms that might be needed to explore new information rather than exploiting already known knowledge based on training data. \\
    & - \textbf{Degree of flexibility in Solution Space}: The solution space needs to be designed in a flexible way to accommodate the adapting changes within the learning/knowledge space.\\
    & - \textbf{Sensitivity Level}: Based on the degree of flexibility in design space, AI-enabled components need to have specific level of sensitivity to prior.  \\
    & - \textbf{Operations}: AI-enabled component capability in using the new information it receives from the interactions with the environment, which can be internal (through the transformation executed by other components in the system) through its interfaces.\\
    & - \textbf{Events}: Information such as location-related data should determine how the AI-enabled component should generate behavior so that the other components in the system can respond to. \\
    & - \textbf{Spatial Information Connectivity}: AI-enabled component's mechanism to transfer the information of changes spatially within the system.\\
    & - \textbf{Type of change}: Possible change states of component, mechanism of change, Sources of uncertainty in changes.\\ 
    & - \textbf{Feedback cycle}: Feedback cycles that AI-enabled component needs in the system through its interfaces to adapt its behavior based on changes in environment.\\ 
    & - \textbf{Interactions}: Internal interfaces of AI-enabled component with other parts of the system and how they filter/transform environment into specific interfaces.  \\
    & - \textbf{Noise handling}: How AI-enabled component's interface filters the noise received from the environment. \\
    & - \textbf{Trigger mechanism}: The Cooperation should be triggered at the interface based on the set of high-level missions.\\
    & - \textbf{Local interaction rules and constraints}: The AI-enabled components start to interact with each other locally based on predefined policy on dispatching, cloning and disposing information.\\
    & - \textbf{Human interaction rules and constraints}: In a case where humans need to interact with AI-enabled component(s), they need to follow specific rules.\\
    & - \textbf{Model Card Report}: 
    Training Algorithms, AI-enabled component’s Reasoning Capability, learning performance metrics, sensitivity level, training data characterization. In general, information reported in the Model Cards.  \\
    & - \textbf{Verification Strategy}:
    The procedures done to Verify abstractability and generality of AI-enabled component. The procedures for Verifying maintainability, changeability, and robustness of AI-enabled component. Agent-based simulation. AI-enabled component test cases of cooperation and interaction.\\
    \hline
    \end{tabular}
\end{table}

\section{Conclusion}

We have presented ongoing work in the development of a template for the interface description of AI-enabled components and listed key features that we suggest should be reported. Interface description templates give system designers a framework to evaluate autonomy performance of different AI-enabled components with a focus on reusability in appropriate contexts when needed. 

The proposed interface description templates should be effective in defining clear boundaries between AI-enabled components and their interfaces with other components and the environment, highlighting the critical aspects of different characteristics of their autonomy. We view the proposed interface description as aiding a more formal procedure for assessing reusability of AI-enabled systems, thus contributing to velocity and effectiveness of development efforts. 

The proposed interface description templates merged common interface descriptions for HW systems and SW systems with autonomy-specific information. This information was derived from the specific challenges that system designers encounter when dealing with AI-enabled capabilities, inspired by how Model Cards support reusability assessments of AI algorithms.

\end{document}